\documentclass[aps,superscriptaddress,floats,showpacs,floatfix,preprint]{revtex4}
\usepackage{bm}
\usepackage{graphicx}
\usepackage{subfigure}
\usepackage{amssymb}
\usepackage{amsmath}
\usepackage{color}
\usepackage[a4paper,left=1.5cm, right=1.3cm, top=3.0cm,bottom=2cm]{geometry}
\begin{document}
\newcommand{\joerg}[1]{\textcolor{red}{#1}}
\newcommand{\janet}[1]{\textcolor{blue}{#1}}

\title{Extreme dissipation event due to plume collision in a turbulent convection cell}
\author{J\"org Schumacher}
\affiliation{Institut f\"ur Thermo- und Fluiddynamik, Technische Universit\"at Ilmenau, Postfach 100565, D-98684 Ilmenau, Germany}
\author{Janet D. Scheel}
\affiliation{Department of Physics, Occidental College, 1600 Campus Road, M21, Los Angeles, California 90041, USA}
\date{\today}

\begin{abstract}
An extreme dissipation event in the bulk of a closed three-dimensional turbulent convection cell is found to be correlated with a strong
reduction of the large-scale circulation flow in the system that happens at the same time as a plume emission event from the bottom plate. 
The reduction in the large-scale circulation opens the possibility for a nearly frontal collision of down- and 
upwelling plumes and the generation of a high-amplitude thermal dissipation layer in the bulk. This collision is locally connected to  
a subsequent high-amplitude energy dissipation event in the form of a strong shear layer. Our analysis illustrates the impact of 
transitions in the large-scale structures on extreme events at the smallest scales of the turbulence, a direct link that is observed 
in a flow with boundary layers. We also show that detection of extreme dissipation events which determine the far-tail statistics of the 
dissipation fields in the bulk requires long-time integrations of the equations of motion over at least hundred convective time units. 
\end{abstract}
\maketitle

\section{Introduction}
The highly nonlinear dynamics of fully developed turbulence generates high-amplitude fluctuations of the flow fields and 
their spatial derivatives. For the latter, amplitudes can exceed the statistical mean values by several orders of magnitude \cite{Yeung2015}.
From a statistical point of view, extreme events correspond to amplitudes in the far tail of the probability density function of the considered 
field. Although the events are typically rare,  they can appear much more frequently than for a Gaussian distributed field -- a manifestation 
of (small-scale) intermittency in turbulence \cite{Frisch1994,Ishihara2009}. From a mathematical perspective, these high-amplitude events are 
solutions of the underlying dynamical equations which display a very rapid temporal variation with respect to a norm defined for the whole fluid 
volume \cite{Doering2009}.  Typical quantities which can be probed are the vorticity  or (local) enstrophy \cite{Lu2008,Donzis2010}, local strain 
\cite{Schumacher2010} or the magnitude of temperature, and passive scalar derivatives \cite{Kushnir2006}. Numerical 
studies of extreme events in turbulence have been performed in cubes with periodic boundaries in all three space dimensions 
\cite{Boratav1994,Donzis2010}.  With increasing Reynolds number these extreme events in box turbulence are concentrated in ever finer 
filaments or layers \cite{Yeung2015}. 
 
Alternatively, extreme dissipation events can be connected to flow structures in wall-bounded flows that have a large spatial coherence 
and exist longer than the typical eddies or plumes. Such dissipation events are observed for example in connection with ramp-cliff structures 
of the temperature \cite{Corrsin1962, Antonia1979}, with superstructures of the velocity \cite{Marusic2010} in atmospheric boundary layers, 
or with very-large scale motion in pipe flows \cite{Hellstroem2015}. High-dissipation events are then detected inside the container as well as at the 
edge of the boundary layers.

In this work, we demonstrate a direct dynamical link between a transition of the large-scale turbulent fields and the rare high-amplitude 
events of the spatial derivatives which are sampled at the smallest scales of the turbulent flow far away from the boundary layers.
The system is a three-dimensional turbulent Rayleigh-B\'{e}nard convection (RBC) flow in a closed cylindrical cell. We show how the formation of a 
rare high-amplitude dissipation rate event in the bulk of the convection cell can be traced back to a plume emission from the bottom plate 
coinciding with a strong fluctuation of the large-scale circulation which exists in closed turbulent flows  \cite{Ahlers2009,Chilla2012}. In the large-scale 
fluctuation event, the large-scale circulation (LSC) roll is significantly weakened and re-oriented afterwards. In the absence of the large-scale ordering 
circulation (which would sweep the plumes along with it), a collision between a hot upwelling and cold downwelling plume is triggered which 
generates strong local gradients. Such extreme dissipation events are very rare in the bulk since most of the 
viscous and thermal dissipation is inside the boundary layers at the top and bottom plates. This has been shown in several direct numerical
simulations  (DNS) of convection \cite{Emran2008,Kaczorowski2013,Scheel2013}. In our five high-resolution spectral element simulations at
different Rayleigh and Prandtl numbers, we monitored the fourth-order moments of the thermal and kinetic energy dissipation rates in the bulk of the 
cell far away from the boundary layers. After finding one data point in one run which was much larger than the rest, we reran 
this full simulation twice in the interval around this extreme event at a monitoring frequency five and fifty times higher in order to analyze the dynamics in detail. 
Our detected rare event reveals a direct connection between a strong large-scale fluctuation of the velocity and a small-scale extreme dissipation (i.e. velocity 
derivative) event,  thus bridging the whole cascade range of the turbulent flow.   

\section{Numerical model}
We solve the three-dimensional Boussinesq equations for turbulent RBC in a cylindrical cell of height $H$ and 
diameter $d$. The equations for the velocity field $u_i(x_j,t)$ and the temperature field $T(x_j,t)$ are given by
\begin{align}
\label{ceq}
\partial_i u_i &=0\,,\\
\label{nseq}
\partial_t u_i +u_j \partial_j u_i &=-\partial_i p+\nu \partial_j^2 u_i+ g \alpha (T-T_0) \delta_{iz}\,,\\
\label{pseq}
\partial_t T +u_j \partial_j  T&=\kappa \partial_j^2 T\,,
\end{align}
with $i,j=x,y,z$ and the Einstein summation convention is used.
The kinematic pressure field is denoted by $p(x_j,t)$ and the reference temperature by $T_0$. The aspect ratio of the convection 
cell is $\Gamma=d/H=1$ with $x,y\in [-0.5,0.5]$ and $z\in [0,1]$. The Prandtl number which relates the kinematic viscosity $\nu$ and 
thermal diffusivity $\kappa$ is 
given by 
\begin{equation}
Pr=\frac{\nu}{\kappa}\,. 
\end{equation}
The Rayleigh number is given by 
\begin{equation}
Ra=\frac{g\alpha\Delta T H^3}{\nu\kappa}\,.
\end{equation}
Here, the variables $g$ and $\alpha$ denote the acceleration due to 
gravity and the thermal expansion coefficient, respectively. The temperature difference between the bottom and top plates is $\Delta T$. 
In a dimensionless form all length scales are expressed in units of $H$, all velocities in units of the free-fall velocity $U_f=\sqrt{g\alpha\Delta T H}$ and 
all temperatures in units of $\Delta T$. Times are measured in units of the convective time unit, the free fall time $T_f=H/U_f$. 

We apply a spectral element method in the present direct numerical simulations (DNS) in order to resolve the gradients of velocity and temperature accurately 
\cite{bib:nek5000}. More details on the numerical scheme and the appropriate grid resolutions can be found in Ref. \cite{Scheel2013}, and resolution
of  higher-order moments of the dissipation rates in \cite{Schumacher2014}.  No-slip boundary conditions are applied for the velocity at all the walls. 
The top and bottom walls are isothermal and the side wall is thermally insulated.

The cylindrical convection cell is covered by $N_e$ spectral elements. On each element all turbulent fields are expanded by $N$th--order 
Lagrangian interpolation polynomials with respect to each spatial direction. Table \ref{Tab1} summarizes our highest Rayleigh number runs on massively 
parallel supercomputer simulations which have been carried out on up to 262144 MPI tasks. In the course of these production runs we conducted an analysis in 
which we searched for extreme dissipation events by means of the fourth-order moments obtained in an inner volume of the closed cylindrical cell.  

The sequence around the extreme dissipation event was rerun twice to generate 
a fine sequence of one hundred snapshots with a separation of 0.143 free fall times $T_f$  and then a very fine sequence of five hundred snapshots 
with a separation of 0.029 $T_f$.
\begin{table}
\begin{center}
\begin{tabular}{ccccc}
\hline\hline
Run & $Ra$ &  $Pr$ & $N_{e}$ & $N$ \\ \hline
1 & $10^8$ & 0.7 & 256,000 & 11 \\
2 & $10^9$ & 0.7 & 875,520 & 11 \\
3 & $10^{10}$ & 0.7 & 2,374,400 & 11 \\
4 & $10^7$  & 0.021 & 875,520 & 11 \\
5 & $10^8$  & 0.021 & 2,374,400 & 13 \\ \hline\hline
\end{tabular}  
\caption{Parameters of the different spectral element simulations. We show the Rayleigh number $Ra$, the Prandtl number $Pr$, 
the total number of spectral elements $N_e$, and the polynomial order $N$ of the Lagrangian interpolation polynomials in each 
of the three space directions.}
\label{Tab1}
\end{center}
\end{table}
\begin{figure}
\begin{center}
\includegraphics[scale=0.7]{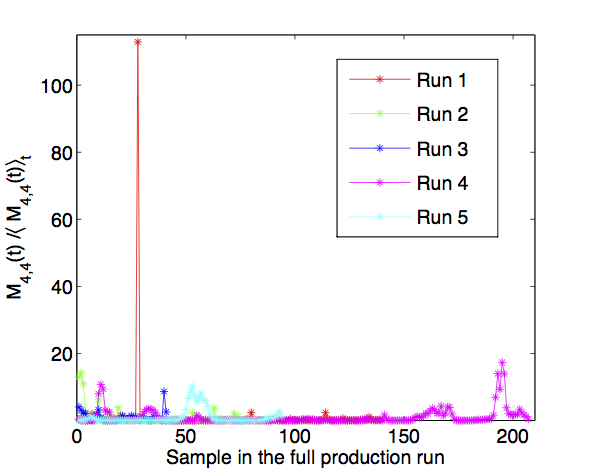}
\caption{(Color online) Appearance of extreme thermal dissipation events in the bulk for five simulation runs
which are listed in Table \ref{Tab1}. The normalized fourth-order thermal dissipation rate moments $M_{4,4}(t)/\langle M_{4,4}(t) \rangle_t$ are shown 
versus the number of statistically independent samples saved in the simulation runs in subvolume $V_4$ which is approximately $V_0/5$.} 
\label{fig1_app}
\end{center}
\end{figure}
\begin{figure}
\begin{center}
\includegraphics[scale=0.45]{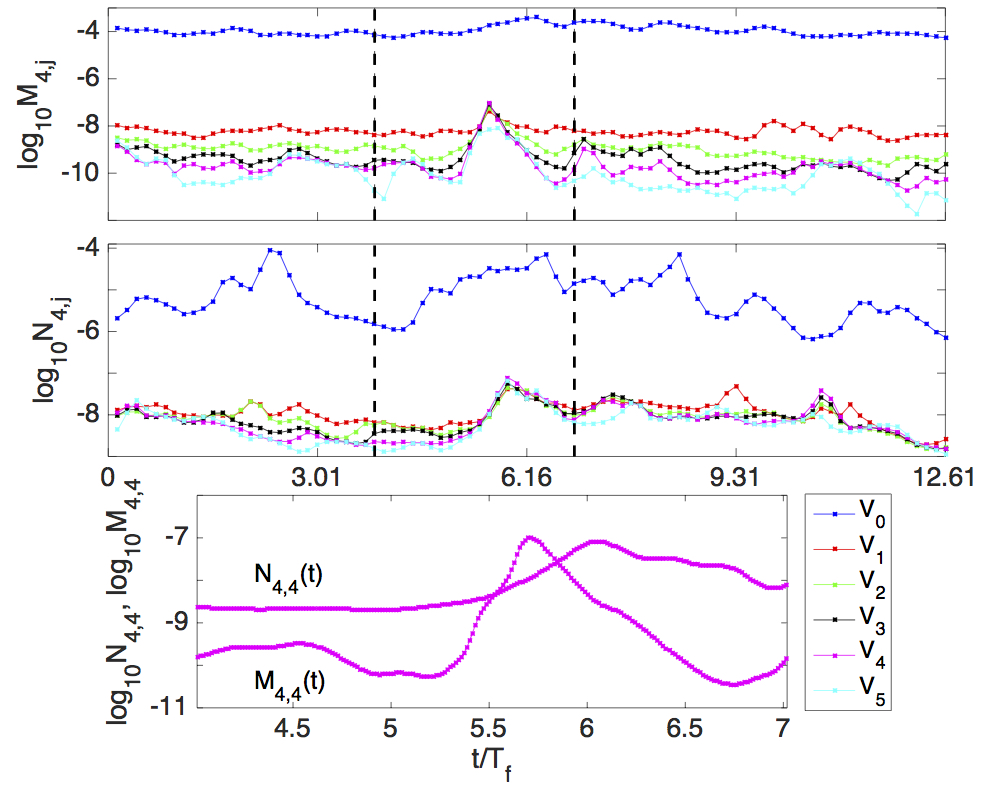}
\caption{(Color online) Monitoring of the evolution of the extreme dissipation event in the bulk by means of the fourth moments of 
the thermal dissipation, $M_{4,j}$, (top panel) and kinetic energy dissipation, $N_{4,j}$, (mid panel). We display the moments in 
six different subvolumes $V_1\dots V_5$ and the whole cell $V_0$. The vicinity of the extreme event is marked by the vertical 
dashed lines and replotted in the bottom panel. These data are taken from the run with the finest temporal resolution.}
\label{fig1}
\end{center}
\end{figure}
\begin{figure}
\begin{center}
\includegraphics[scale=0.45]{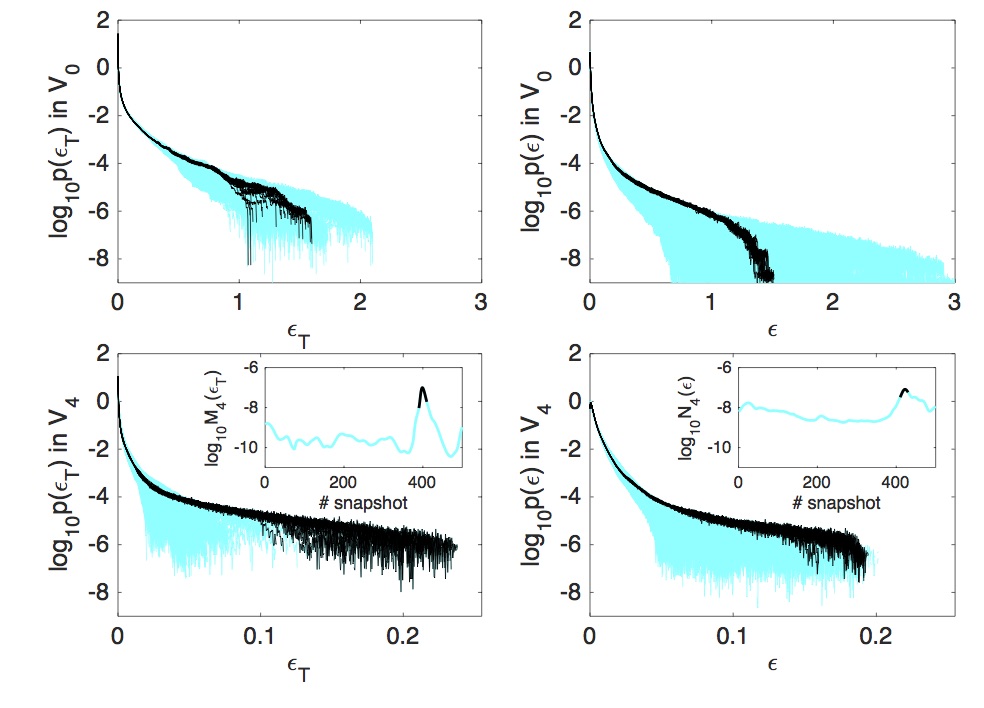}
\caption{(Color online) 
Five hundred individual probability density functions (PDFs) of the thermal dissipation rate $\epsilon_T$ in the left column and 
of the kinetic energy dissipation rate $\epsilon$ in the right column which are obtained from the run with the very fine time resolution. 
Data are for run 1. The insets replot data from the bottom panel of figure \ref{fig1}. The data in the vicinity of the local maxima are always
highlighted as dark curves.}
\label{fig2_app}
\end{center}
\end{figure}
\begin{figure}
\begin{center}
\includegraphics[scale=0.16]{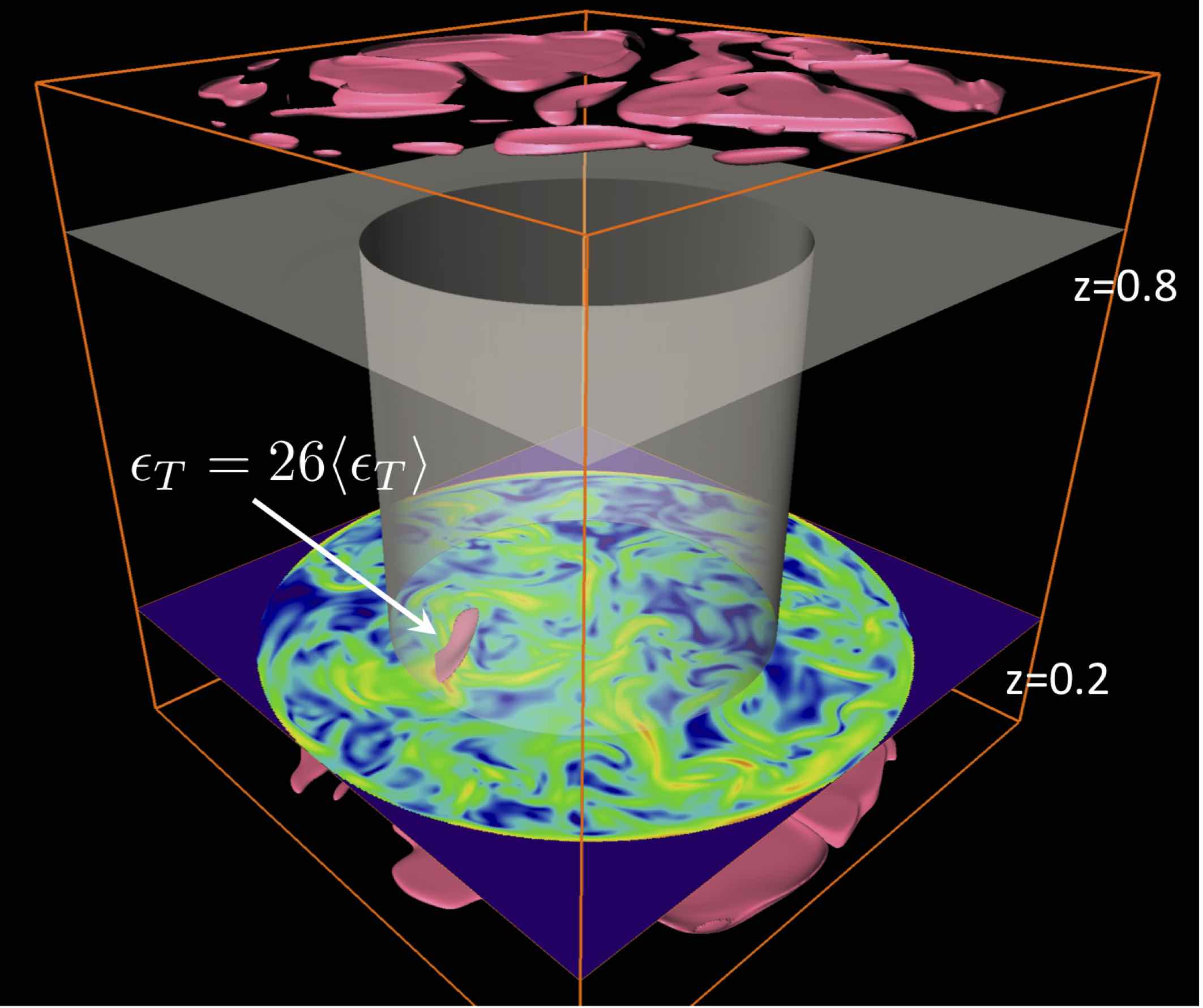}
\caption{(Color online) Extreme thermal dissipation event in the bulk at time $T_{\ast}=5.73$.  Combined plot of thermal 
dissipation rate (isosurfaces at $26 \langle\epsilon_T\rangle_{V_0,t}$) and kinetic energy dissipation rate (horizontal contour 
slice) on a logarithmic scale. Contour slice levels are from blue ($\log_{10}\epsilon  \le -4$) to red ($0.9 \le \log_{10}\epsilon)$. 
The inner cylinder stands for subvolume $V_4$ with $r\le 0.3$ and $0.2\le z\le 0.8$.}  
\label{fig2}
\end{center}
\end{figure}

\section{Results}
\subsection{Detection by fourth-order moments} 
The starting point of the analysis is the time evolution of the fourth-order moments of the kinetic energy dissipation rate, 
\begin{equation}
\epsilon(x,y,z,t)=2\nu S_{ij}S_{ji}\,,
\end{equation} 
with $S_{ij}=(\partial_i u_j+\partial_j u_i)/2$,  and the thermal dissipation rate, 
\begin{equation}
\epsilon_T(x,y,z,t)=\kappa G_i^2\,,
\end{equation} 
with $G_i=\partial_i T$. The Rayleigh-B\'{e}nard flow in the cylindrical cell 
obeys statistical homogeneity in the azimuthal direction only. All statistics will  therefore depend on the size 
of the sample volume. We have monitored the moments in six successively smaller cylindrical 
subvolumes which are nested in each other. We define $r_0=0.5>r_1=0.45>\dots>r_5=0.25$ and $h_0=1>h_1>\dots >h_5=0.5$ and 
$V_j= \{ (r,\phi,z)\, |\,r\le r_j\;,(1-h_j)/2\le z\le (1+h_j)/2 \}$ with $j=0\dots 5$. The volume $V_0$ is the full cell.  
Fourth-order moments of both dissipation rates are given by
\begin{equation}
M_{4,j}(t)=\langle \epsilon_T^4\rangle_{V_j}\quad\mbox{and}\quad N_{4,j}(t)=\langle \epsilon^4\rangle_{V_j}\,.
\end{equation} 

In figure \ref{fig1_app} the normalized moments of the thermal dissipation rate are shown for five different runs which are obtained
at the highest Rayleigh numbers and two different Prandtl numbers (see table \ref{Tab1}). In the primary production runs we analysed 
the kinetic energy and thermal dissipation rate in the subvolume $V_4$ that is sufficiently far away from all boundaries. Since 
the simulation runs have a different number of time step widths and a different number of data output steps, the moments are shown versus
the number of samples. It is clearly visible that in all runs the volume averages can go far beyond the means at certain times. However, the strongest outlier is 
observed for run 1 at $Ra=10^8$ and $Pr=0.7$. Therefore, the discussion in this work is dedicated to run 1.

In figure \ref{fig1} we display $M_{4,j}(t)$ (top panel) and $N_{4,j}(t)$ (mid panel) on a semi-logarithmic plot for run 1. 
Data are obtained over a time interval with an output of one hundred snapshots separated by 0.143 free 
fall times units (see top and mid panels of the figure). $M_{4,0}(t)$ remains nearly unchanged and $N_{4,0}(t)$ fluctuates more strongly, 
but there is no large event that stands out. 
The reason is that a major part of both the thermal variance and of the kinetic energy is dissipated in the boundary layers
of the temperature and velocity fields close to the walls, respectively \cite{Emran2008,Kaczorowski2013,Scheel2013}. Only 
in the successively smaller subvolumes $V_j$, that are
nested increasingly deeper in the bulk, is the extreme bulk dissipation event detected by the corresponding fourth order moment. 
It is seen that $M_{4,4}(t)$  grows by three orders of magnitude within $T_f/2$. The bottom panel of figure \ref{fig1} shows that a local, 
but less strong maximum of $N_{4,4}(t)$ occurs approximately $T_f/2$ after the peak in $M_{4,4}$. 

The significance of this event for the small-scale statistics of the temperature and velocity derivatives in the bulk region is demonstrated 
in figure \ref{fig2_app}.  In both cases the fattest tail corresponds with this high-amplitude event as seen in the bottom panels of figure \ref{fig2_app}. 
We display five hundred individual probability density functions (PDFs), each taken at one instant in time. These data have been obtained in a repetition 
run at the highest temporal resolution in order to resolve the event better. The vicinity of the high-dissipation event is colored differently in both dissipation 
rates. It is also seen that the high-dissipation bulk event does not contribute significantly to the far tails of the PDFs when averaged over the whole 
convection cell including all boundary layers. The resulting extension of the far tail of the time-averaged PDFs in the bulk was already shown in 
ref. \cite{Scheel2013}.
\begin{figure}
\begin{center}
\includegraphics[scale=0.5]{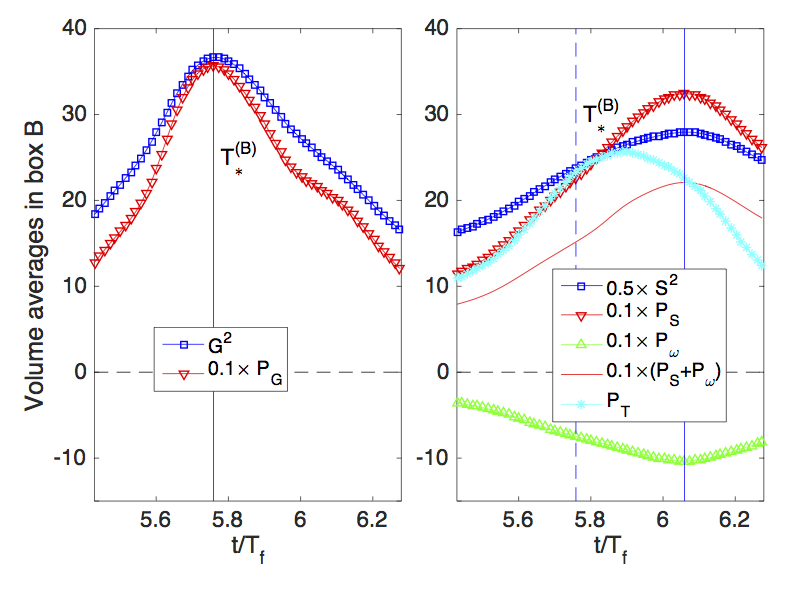}
\caption{(Color online) Time evolution of production terms and magnitudes in the course of the extreme event. All quantities are now volume averages taken for 
box ${\cal B}$. The maximum of $G^2$ is at $T_{\ast}^{({\cal B})}=5.759$ and is slightly shifted with respect to $T_*$ in $V_4$ because 
$V_4 \gg {\cal B}$ and the temporal resolution is finer. Left:  temperature gradient square and production term $P_G$ (see eq. (\ref{gradbal})). Right: local strain and production terms $P_S$,
$P_{\omega}$ as well as $P_T$ (see eq. (\ref{strainbal})). The peaks of $G^2$ (left) and $S^2$ (right) are indicated by solid vertical lines. The dashed 
vertical line in the right panel is the maximum of $G^2$. Terms are partly rescaled as indicated in the legend.}
\label{fig3}
\end{center}
\end{figure}

\subsection{Link between high-amplitude thermal and kinetic energy dissipation events} 
Figure \ref{fig2} shows isosurfaces of the thermal dissipation rate at $\epsilon_T= 26 \langle\epsilon_T\rangle_{V_0,t}$ 
which are mostly found close to the top and bottom plates. The same holds for kinetic energy dissipation, but is not shown. It is the high-thermal-dissipation 
sheet at $T_{\ast}=5.73$ which is mostly inside $V_4$ that contributes to the local maxima of $M_{4,j}(t)$ for $j>0$ in figure \ref{fig1}. It can be also 
seen that the local maximum of $\epsilon_T(x,y,z,t)$ coincides with a local maximum of $\epsilon(x,y,z,t)$. The temperature front generates 
a strong shear layer which is manifest as a delayed high-amplitude energy dissipation event. 

We refined the analysis, both in space and time. We zoom into 
the small box ${\cal B}=\{(x,y,z)\in[-0.11,-0.05]\times[-0.34,-0.14]\times[0.15,0.33]\}$ that encloses the high-amplitude thermal dissipation layer.
The balance equation for the square of the magnitude of $G_i$ is given by \cite{Pumir1994,Brethouwer2003}
\begin{equation}
\frac{\mbox{d}G^2}{\mbox{d}t}=-2G_i S_{ij} G_j +2\kappa G_i \frac{\partial^2 G_i}{\partial x_j^2}\,.
\label{gradbal}
\end{equation}
The first term on the right hand side is the gradient production term, $P_G$. Local shear strength is measured by the  square 
of the magnitude of the rate of strain tensor $S^2=S_{ij} S_{ji}$. The balance equation for $S^2$ (see also \cite{Holzner2008}) 
has to be extended by a temperature production term and is given by
\begin{align}
\frac{\mbox{d}S^2}{\mbox{d}t}=&-2S_{ij} S_{jk} S_{ki}-\frac{1}{2} \omega_i S_{ij} \omega_j -2 S_{ij} \frac{\partial^2 p}{\partial x_i \partial x_j} \nonumber\\
                                                &+2\nu S_{ij}\frac{\partial^2 S_{ij}}{\partial x_k^2} + 2g\alpha S_{zi}G_i\,.
\label{strainbal}                                                
\end{align}
We have three production terms: strain production (1st, $P_S$), enstrophy consumption (2nd, $P_{\omega}$) and production due to coupling 
to the temperature gradient (last, $P_T$).  
\begin{figure}
\begin{center}
\vspace{0.2cm}
\includegraphics[scale=0.13]{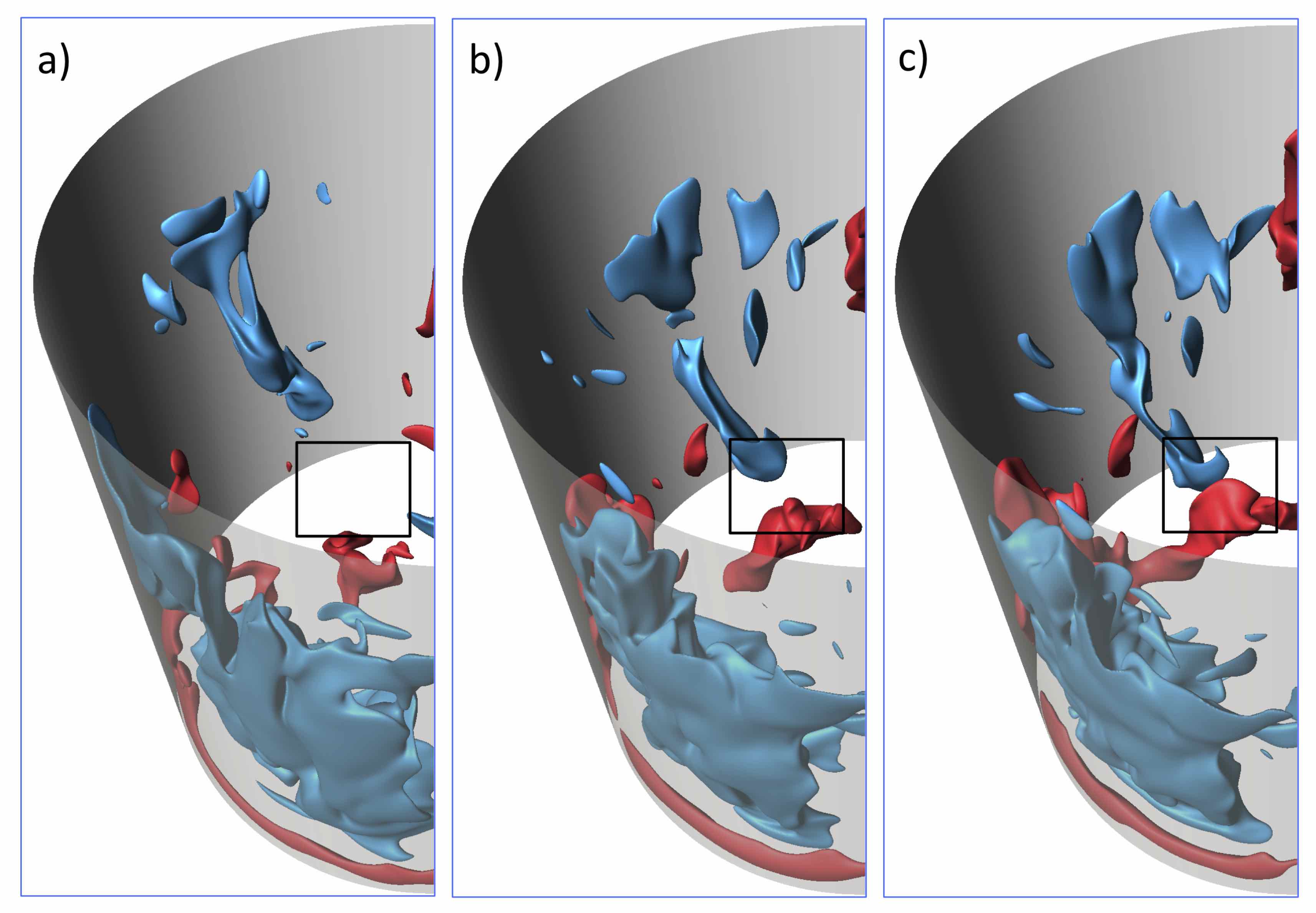}
\caption{(Color online) Isocontour plot of the vertical convective current for times: (a) $T_{\ast}-1.146$, (b) $T_{\ast}-0.296$ and (c) $T_{\ast}$.
Blue is for downwelling at $\sqrt{Ra Pr}\,u_z T=-900$, red for upwelling plumes at $\sqrt{Ra Pr}\,u_z T=1000$. 
The collision region is indicated by a box.}  
\label{fig4}
\end{center}
\end{figure}
Figure \ref{fig3} displays the time evolution of volume averages over ${\cal B}$ for both gradient magnitudes and the corresponding production terms 
in eqns. (\ref{gradbal}) and (\ref{strainbal}), respectively. The maximum of $G^2$ coincides with the one of $P_G$ (see left panel). 
The same holds for the maximum of $S^2$ and the ones of $P_S$ and $\left|P_{\omega}\right|$, respectively (right panel). 
We also confirm that $\max \langle S^2\rangle_{\cal B}$  lags behind $\max \langle G^2\rangle_{\cal B}$ (see also figure \ref{fig1}), a 
result which is also robust for different sizes of ${\cal B}$. The time of maximum production 
by $P_T$, falls right between those for $P_G$ and $P_S+P_{\omega}$. This shows that the temperature gradient occurs first, followed by 
strong shear generation since the colliding fluid masses have to move around each other.   

\subsection{Formation of colliding plumes} 
How is the high-ampli\-tude thermal dissipation layer formed? Figure \ref{fig4} plots isosurfaces 
of the vertical component of 
the convective heat current vector  $j^c_z=\sqrt{Ra Pr}\,u_z T$ at three instants. Since $0\le T\le 1$, a negative isolevel of $j_z^c$ corresponds 
to a downwelling and a positive one to an upwelling plume. The box in the panels indicates the collision point of two plumes in the bulk at time 
$T_{\ast}$. This collision is caused by the large hot plume from the bottom and a second extended cold plume that falls down at the side wall and turns 
into the bulk. The high-amplitude thermal dissipation layer is formed at the collision site. The event is comparable with rapid growth events of 
enstrophy in box turbulence \cite{Lu2008}. There colliding vortex rings maximized enstrophy growth. Our nearly frontal plume collision can be 
considered thus a rare event and appears in three-dimensional convection flow much less frequently than in two-dimensional ones \cite{Chandra2013}.

\begin{figure}
\begin{center}
\includegraphics[scale=0.15]{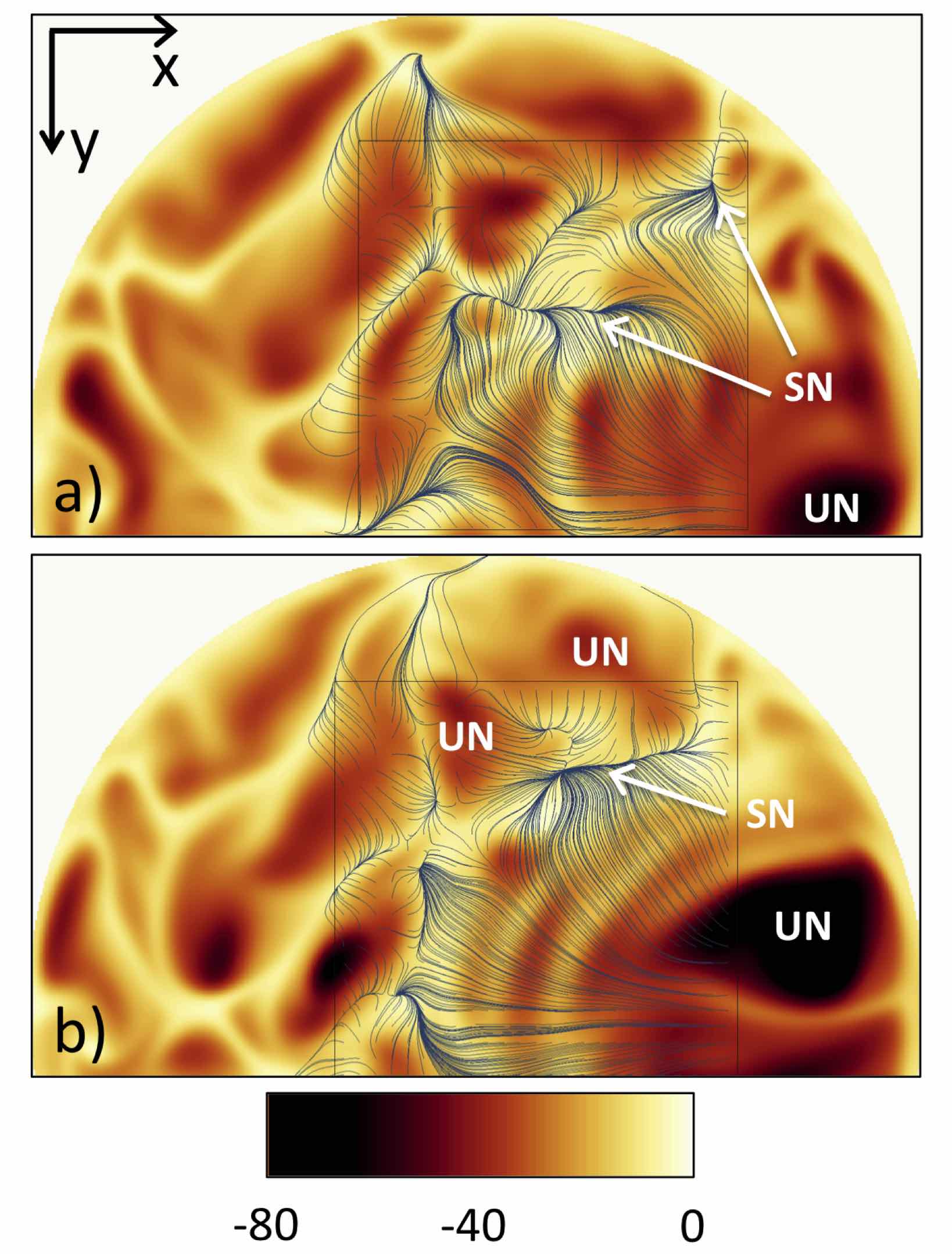}
\caption{(Color online) Strong plume formation at the bottom plate. Contour plots of $\partial T/\partial z$ at $z=0$ are shown together 
with field lines of the skin friction field $(\partial u_x/\partial z, \partial u_y/\partial z)$. Times are $T_{\ast}-1.146$ for (a) and
$T_{\ast}-0.296$ for (b). For better visibility, we seed the skin friction lines only in a square box around the rising plume. 
Stable nodes (SN) and unstable nodes (UN) are indicated.}
\label{fig5}
\end{center}
\end{figure}

First we will investigate the rising hot plume. Figure \ref{fig5} displays contours of $\partial T/\partial z$ at the bottom plate.
Local maxima are indicators for rising plumes \cite{Bandaru2015}. On top of contours we plot field lines of the skin friction field which is 
given by $\partial_i u_j|_{z=0}=(\partial_z u_x, \partial_z u_y)$ \cite{Chong2012}. Locally downwelling fluid impacts the bottom plate and generates 
unstable node points (UN) of the skin friction field. Skin friction lines, which arise from these nodes, form a strong front which starts to form in
panel (a) and is moved ``upward'' in panel (b) of figure \ref{fig5}. Saddle points (SP) or stable nodes (SN) are formed between the unstable 
nodes. The unstable manifold of a saddle \cite{Bandaru2015} or a sequence of stable nodes, as being the case here, are the preferred sites 
of plume formation. It is the persistence and convergence of these critical points for a certain time span which causes the rise of a large plume 
from the bottom just before $T_{\ast}$, that then collides with the downwelling cold plume at $T_{\ast}$.

\subsection{Plume collision due to transition of large-scale flow} 
This raises the last question, namely does a change in the large-scale dynamics enable such a rare 
plume collision event? It is well-known that in  closed convection cells a large-scale circulation (LSC) exists 
\cite{Ahlers2009,Chilla2012}. In cells with $\Gamma=1$, the LSC consists of one big roll which forces the 
plumes to move along the top or bottom plate, and then to rise dominantly on one side of the cell and to fall down 
on the other side. This ordering influence stops when the large-scale circulation decelerates strongly and becomes 
re-oriented. Such events have been studied statistically in experiments \cite{Sreenivasan2002, Brown2006,Xi2007}, 
numerically in two-dimensional  \cite{Chandra2013,Petschel2011,Poel2012} or three-dimensional \cite{Mishra2011} 
convection as well as in low-dimensional models \cite{Brown2008}.

\begin{figure}
\begin{center}
\vspace{0.2cm}
\includegraphics[scale=0.14]{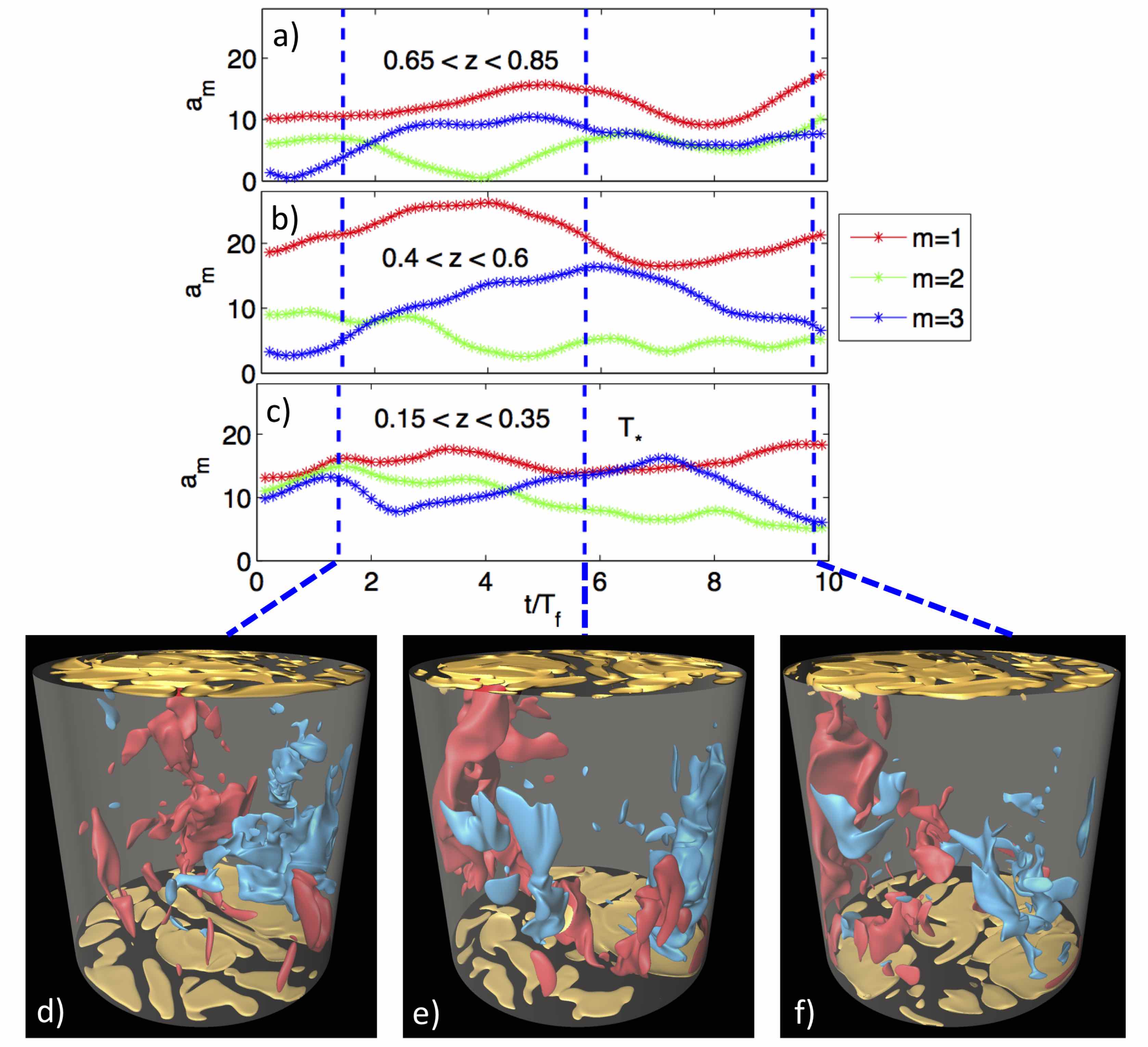}
\caption{(Color online) Time evolution of the three largest Fourier mode amplitudes obtained for $\overline{u_zT}(\phi,t)$.
(a) $z_1=0.65< z< z_2=0.85$. (b) $0.4<z<0.6$. (c) $0.15<z<0.35$. Points $0.4< r < 0.48$ were taken in radial 
direction. At the bottom we add three snapshots of the convective current $\sqrt{Ra Pr}\,u_z T$ (blue for level of -900 and red for 
level of 1100) together with isocontours of $\epsilon_T= 0.1 \approx 26\langle\epsilon_T\rangle_{V_0,t}$. (d) $t=1.43$. (e) $t=T_{\ast}=5.73$. (f) $t=9.74$.}
\label{fig6}
\end{center}
\end{figure}

We quantified the large-scale dynamics by taking a spatial average with respect to the radial and vertical coordinates. 
We define 
\begin{equation}
\overline{u_zT}(\phi,t)=\frac{1}{{\cal V}_r}\int_{r_1}^{r_2}\int_{z_1}^{z_2} u_zT(r,\phi,z,t)\, rdrdz\,,
\label{average}
\end{equation}
with ${\cal V}_r=\pi(r_2^2-r_1^2)(z_2-z_1)$. The complex three-dimensional structure of the up- and downwelling
convective currents in the closed cell  is thus reduced to a one-dimensional signal. The locally averaged convective current $\overline{u_zT}(\phi,t)$ 
is expanded in a Fourier series for each instant
\begin{equation}
\overline{u_zT}(\phi,t) = \sum_{m=1}^N a_m(t)\cos(m\phi+\gamma_m(t))\,.
\label{average1}
\end{equation}
Figure \ref{fig6} displays the amplitude of the first three modes, $a_1(t)$ to $a_3(t)$. We have chosen different vertical 
intervals $[z_1,z_2]$, in the upper and lower sections of the cell as well as in the center. At the beginning of the time window, 
we find $a_1>a_2>a_3$ in all sections of cell. This indicates that a one-roll circulation pattern dominates the LSC as is supported by 
the isocontours in Figure \ref{fig6}(d). In Figures \ref{fig6}(a)--(c),  $a_1$ steadily decreases towards $t=T_{\ast}$ with $T_{\ast}$ 
being the time of the extreme dissipation event. The ratio of the Fourier coefficients is changed to $a_1\sim a_3 > a_2$ for the lower 
section of the cell (see Figure \ref{fig6}(c)), while in the mid 
and upper sections (see Figures \ref{fig6}(a,b)), $a_1>a_3>a_2$ is observed. The growth of the $m=3$ mode demonstrates that up- and downwelling 
convective currents are found now close to each other, in particular in the lower section of the cell, as can be seen by the isocontours in 
Figure \ref{fig6}(e). For $t>T_{\ast}$, we observe a re-establishment 
of the one-roll pattern, as supported by Figure \ref{fig6}(f). The whole process proceeds within $10 T_f$.  Also plotted in Figures \ref{fig6}(d)--(f) 
are the isocontours for large $\epsilon_T$, which always are located near the bottom and top plates. However, in Figure \ref{fig6}e, one sees a 
region of large $\epsilon_T$ in between the upwelling hot and downwelling cold plumes as they collide, consistent with the increase in $\epsilon_T$ 
in the bulk seen in Figure \ref{fig1}.

\section{Summary}
We have connected a far-tail, extreme dissipation event at the small scales in the bulk of a three-dimensional 
Rayleigh-B\'{e}nard convection flow in a closed cell to a reduction event in the LSC accompanied by a plume emission from the bottom boundary 
layer. Such an event is very rare. In five different  simulations spanning a range of $Ra$ and $Pr$ over long evolution times
it was the only very high dissipation event in the bulk away from boundary layers as shown in figure \ref{fig1_app}. 

The detection 
was possible by monitoring the well-resolved fourth-order dissipation moments in the bulk of the cell during the simulations. We also have showed how a 
transition of the large-scale flow structures in the cell can impact the dynamics at the smallest scales, the scales across which the steepest 
gradients are formed. The two events are thus directly linked and bridge the whole scale range of the turbulent cascade. The large-scale 
coherent fluid motion is established here due to the presence of walls which enclose the convection cell. It can be expected that it would be absent 
in box turbulence with periodic boundary conditions.

How frequently does such a high-dissipation event appear? If one takes a typical far-tail amplitude of the 
PDF of $\epsilon_T$ (see lower left panel of figure \ref{fig2_app}) of $p(\epsilon_T)\sim 10^{-6}$ and multiplies it with the binwidth $\Delta \epsilon_T=0.0004$, one 
gets an estimate of the probability of the appearance of a high-thermal-dissipation event in the bulk of $w\approx p(\epsilon_T)\Delta\epsilon_T\approx 4\times 10^{-10}$, i.e., one out of 2.5 billion data points. The bulk volume $V_4$ contains about a fifth of the total cell volume and about 10 per cent of the total number of mesh cells which translates to roughly 40 million cells for $V_4$. That means that one picks such high-dissipation events every 60 to 70 $T_f$ if one continues with the same sampling frequency as in the original production run. Our total integration time for run 1 was 104 $T_f$. Consequently, if one wants to have a complete picture of the small-scale statistics of a wall-bounded turbulent flow then these events have to be incorporated. As our estimate shows, this requires very long-time integrations of the fully resolved Boussinesq equations which becomes increasingly expensive as the Rayleigh number grows or the Prandtl number decreases.

\acknowledgements 
Computing resources have been provided by the John von Neumann Institute for Computing at the J\"ulich Supercomputing 
Centre by Grant HIL09 on Blue Gene/Q JUQUEEN and by Grant SBDA003 of the Scientific Big Data Analytics (SBDA) Program  
on the J\"ulich Exascale Cluster Architecture (JURECA), respectively. We thank F. Janetzko for his support in the SBDA project.

\end{document}